\documentclass[aps,twocolumn,nofootinbib,superscriptaddress]{revtex4-1}

\usepackage[utf8]{inputenc}
\usepackage{amsmath}
\usepackage{amssymb}
\usepackage{latexsym}
\usepackage{array}

\usepackage[usenames,dvipsnames,svgnames,table]{xcolor}
\usepackage{graphicx}
\usepackage{subfig}
\usepackage{wrapfig}
\usepackage{rotating}

\usepackage{dcolumn}
\usepackage{needspace}
\usepackage{enumitem}
\usepackage[tight,nice]{units}
\usepackage{textcomp}
\usepackage{textcomp}
\usepackage{gensymb} 
\usepackage[Symbol]{upgreek} 

\usepackage{tabu}
\usepackage{tabularx}
\usepackage{tabulary}
\usepackage{multirow}

\begin{document}

\title{A simple method of coil design}
\date{\today}

\newcommand{\ETHaffilation}{ETH Zürich, Institute for Particle Physics, CH-8093 Zürich, Switzerland}

\author{M.~Rawlik}
\email{\texttt{mrawlik@phys.ethz.ch}}
\affiliation{\ETHaffilation}

\author{C.~Crawford}
\affiliation{University of Kentucky, Lexington, USA}

\author{A.~Eggenberger}
\affiliation{\ETHaffilation}

\author{K.~Kirch}
\email{\texttt{klaus.kirch@phys.ethz.ch}}
\affiliation{\ETHaffilation}
\affiliation{Paul Scherrer Institute, Villigen, Switzerland}

\author{J.~Krempel}
\affiliation{\ETHaffilation}

\author{F.~M.~Piegsa}
\affiliation{University of Bern, Albert Einstein Center for Fundamental Physics, CH-3012 Bern, Switzerland}
\affiliation{\ETHaffilation}

\author{G.~Quéméner}
\affiliation{Laboratoire de Physique Corpusculaire de Caen, France}

\begin{abstract}
  In this article we present a method to design a coil producing an arbitrarily shaped magnetic field by restricting the path of the coil's wires to a regular grid. The solution is then found by a simple least squares minimum. We discuss practical applications, in particular in the active magnetic field stabilization system of the neutron electric dipole moment experiment at the Paul Scherrer Institute in Villigen, Switzerland. We also publish the software implementation of the method.
\end{abstract}

\maketitle

\section{Introduction}
How to design a coil, or more generally, an arrangement of coils, producing a desired magnetic field? The problem is surprisingly hard and the solutions, how the wire making up the coil should be laid, complicated. The most widespread application of high--performance coils is Magnetic Resonance Imaging (MRI), where gradient coils give the possibility to produce spatial images. Already in the 1980's elaborate methods of MRI coil design have been developed. They range from optimizing positions of discrete windings, where use is made of symmetries specific to MRI, to analytical methods yielding surface current density, which is then discretized. A general overview can be found in \cite{Turner1993}. Another field known for complex, precise coils is plasma--confinement, in particular stellarators \cite{Beidler1990}. There analytical solutions for the surface current density find their use, too.

We would like to present a method that may not be competitive when it comes to precision, but is distinct in its simplicity, also when it comes to construction of its designs. It relies on an algebraic representation of the problem, where coil design is simplified to a simple linear least squares problem. In our method the coils are restricted to a user--defined mesh, making it easy to deal with spatial constraints.


The method has originally been developed to design coils of the active magnetic field stabilization system of the neutron electric dipole moment (nEDM) experiment at the Paul Scherrer Institute in Villigen, Switzerland. This application, described in detail in \cite{Afach2014}, which we discuss later in the paper, requires large (6--8 meters side length) coils. In the presented method the coil systems are designed on a predefined grid. This makes the construction of even complicated coils feasible, despite the size.








We begin with a description of our model in a restricted 2-dimensional case and generalize it to three dimensions. We then present how the model is used to design a coil, based on an example. Further we discuss possibilities of simplifying the solution. Another section is devoted to practical considerations, significant for the eventual construction. Finally, we analyze the design method in the particular case of the magnetic field stabilization system of the nEDM experiment at the Paul Scherrer Institute.

\section{Coils as a linear space}
Consider all possible coils that can be constructed by laying a wire on a surface of a square. The possibilities are endless. Speaking more precisely, as the wires may be shifted by arbitrarily small distances, as they overlap and cross, the problem has inherently an infinite number of degrees of freedom. We present an algebraic representation that reduces the number of degrees of freedom to just a few.

\begin{figure}
  \centering
  \includegraphics[width=0.8\linewidth]{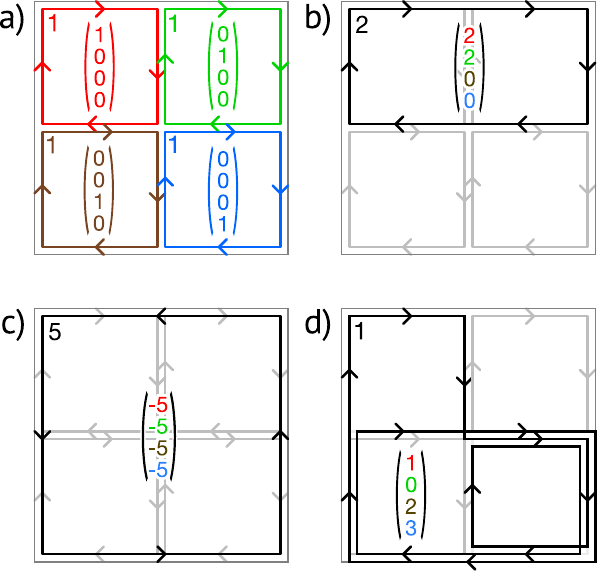}
  \caption{a) A basis of four tile coils on a flat square. Any coil which has its wires restricted to lie on the $2 \times 2$ grid can be represented as a linear combination of the four base tile coils. b, c, d) Three coils are presented together with their explicit coordinates in the basis.}
  \label{fig:coils_tile_basis}
\end{figure}

Let us imagine a square loop of a wire with a current flowing through it -- a coil. It produces a certain magnetic field in the entire space $\mathbf{B}(\mathbf{x})$, which can be easily calculated using the Biot-Savart law (a ready solution can be found e.g.\ in \cite{Reta-Hernandez1998}). By changing the current in the coil we can alter only one parameter of the magnetic field -- the magnitude, but not its shape. It can therefore be said that one coil spans a one--dimensional space of magnetic fields it can produce. Adding a second, different coil creates a system spanning a two--dimensional space of fields, as the magnetic field is additive. Going a step further, four square coils tiled to form a larger square form a four--dimensional space, as shown in Fig.\,\ref{fig:coils_tile_basis}. Any coil restricted to the $2 \times 2$ grid can be represented in the base of the four tile--coils.

The range of magnetic field reachable by coils restricted to a grid is a subset of all possible fields that can be created with coils constructed on the square's surface. The size of the subset is controlled by $N$, the number of tile--coils forming the grid.
In this system a coil is fully described by a vector of $N$ currents, one in each of the tile--coils, denoted by $\mathbb{I}$. The problem of coil design is thereby simplified to finding a vector $\mathbb{I}$.

\begin{figure}
  \centering
  \includegraphics[width=0.5\linewidth]{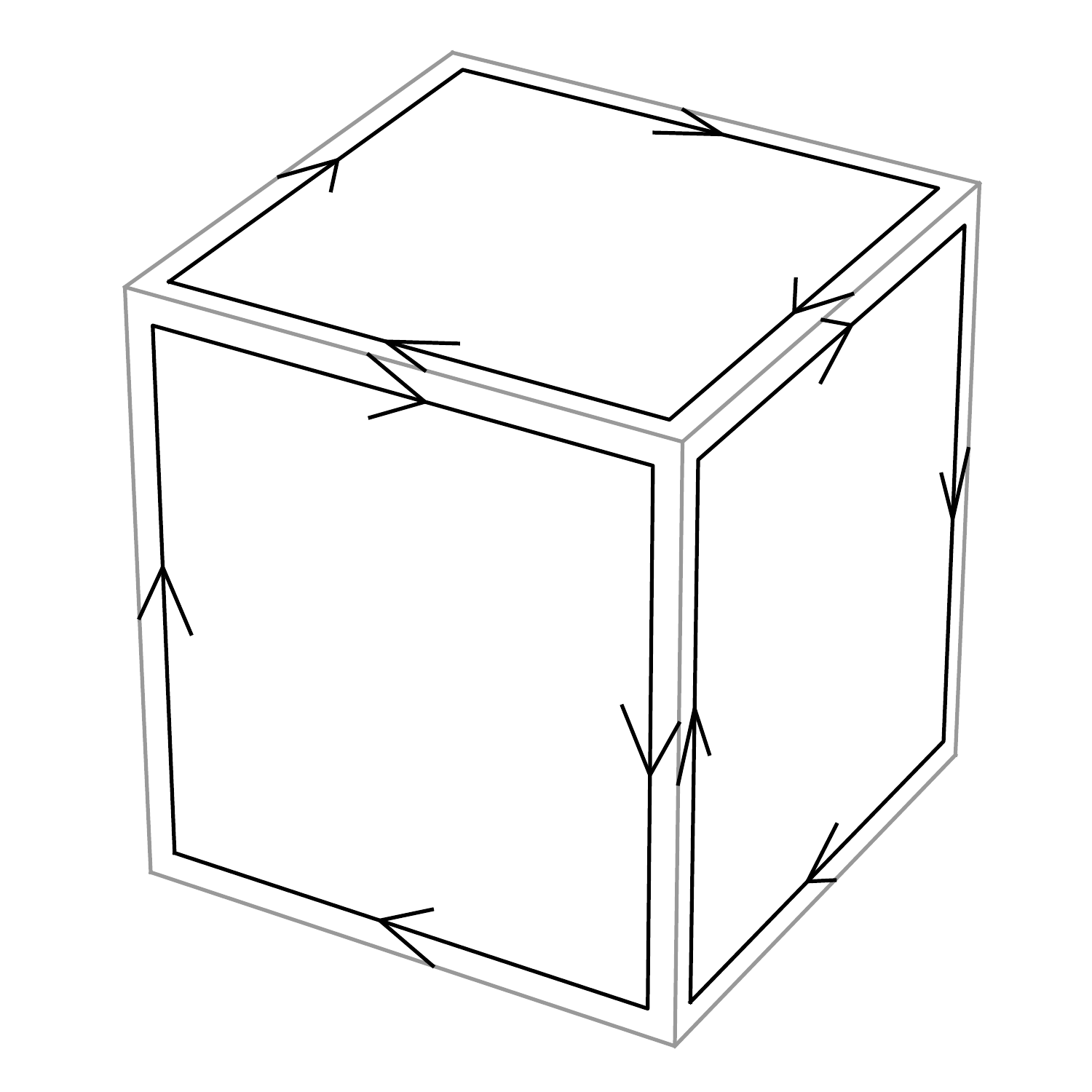}
  \caption{An arrangement of $N = 6$ tile coils on a cube which produces no magnetic field. The currents in the tiles are equal and flow in the directions as indicated. The currents on the invisible faces are analogous to the ones seen in front. For clarity, the coils are depicted slightly smaller; in the model the currents are identical with the edges of the cube.}
  \label{fig:coils_tile_kernel}
\end{figure}

Generalisation onto a cube is simple, a cube being made up of six square faces. Interestingly, in the assembly in a three--dimensional space one degree of freedom is lost.  Figure~\ref{fig:coils_tile_kernel} illustrates, in the simplest case $N = 6$, a configuration in which finite currents in all six coils cancel and no magnetic field is produced. Such a combination of currents can be added to any solution with no effect on the produced field. Effectively, the space of the fields they can produce has dimension 5 ($N-1$). In other words, the mapping of $\mathbb{I}$ onto fields $\mathbf{B}(\mathbf{x})$ has in this case a one--dimensional kernel. This fact is of importance when it comes to numerically solving the system.

This is the foundation of the method. We restrict ourselves to a grid on a cuboid, but in return we can fully describe all coils in the restricted space by a vector of $N$ numbers.

\section{Coil design}
In the problem of coil design one wants to create a coil, or an arrangement of coils, which best approximates a given field in a certain volume, which we will call \emph{the volume of interest}. Rather than considering the whole volume, we pick an ensemble of $m$, points of interest on its surface (the surface is sufficient because $\nabla \cdot \mathbf{B} = 0$). Hence, we look at the magnetic field $\mathbf{B}(\mathbf{x})$ only at these points and gather the values $\mathbf{B}(\mathbf{x}_i)$ for $i = 1 .. m$ into a vector of dimension $3m$ ($B_x$, $B_y$ and $B_z$ in each point), which we shall denote $\mathbb{B}$.


As mentioned before, the magnetic field produced by a coil at any given point in space is proportional to the current in this coil. With many coils present it is a linear combination of the currents of all coils in the system. In absence of an external magnetic field the system of $N$ tiles and $m$ points of interest is thus described by a simple linear equation:
\begin{equation}
  \label{eq:matrix}
  \mathbb{B} = \mathbb{M} \, \mathbb{I}
\end{equation}
where $\mathbb{M} \in \mathbb{R}^{3 m} \times \mathbb{R}^{N}$ is a matrix of proportionality constants. For example, the element $\mathbb{M}_{(5, 2)}$ is the proportionality constant between the current in the second of $N$ coils and the magnetic field in the $y$ direction in the second of $m$ points of interest, $B_x(\mathbf{x}_1)$. The matrix $\mathbb{M}$ can be calculated analytically using the Biot--Savart law.

Equation~(\ref{eq:matrix}), for $3m > N - 1$, is an over--determined set of linear equations, $\mathbb{I}$ being the vector of unknowns. The optimal solution $\mathbb{I}_0$ to produce a $\mathbf{B}_0(\mathbf{x})$ in the volume of interest is found by the ordinary least--squares method:
\begin{equation}
  \label{eq:requirement}
  \mathbb{I}_0 = \mathrm{arg}\,\min_{\mathbb{I}} \left( \mathbb{M} \mathbb{I} - \mathbb{B}_0 \right)^2
\end{equation}
 Depending on the properties of $\mathbb{M}$ the optimum may be multidimensional. In particular, as already mentioned, an arrangement of coils on a cube has a one--dimensional kernel, which will always cause the optimum to be at least a one--dimensional. In these cases we will call $\mathbb{I}_0$ the unique least norm solution, which minimizes the total current in the system. $\mathbb{I}_0$ is the vector of the optimal currents in the tile arrangement of coils for approximating $\mathbf{B}_0(\mathbf{x})$ in the volume of interest.

\begin{figure}
  \centering
  \includegraphics[width=\linewidth]{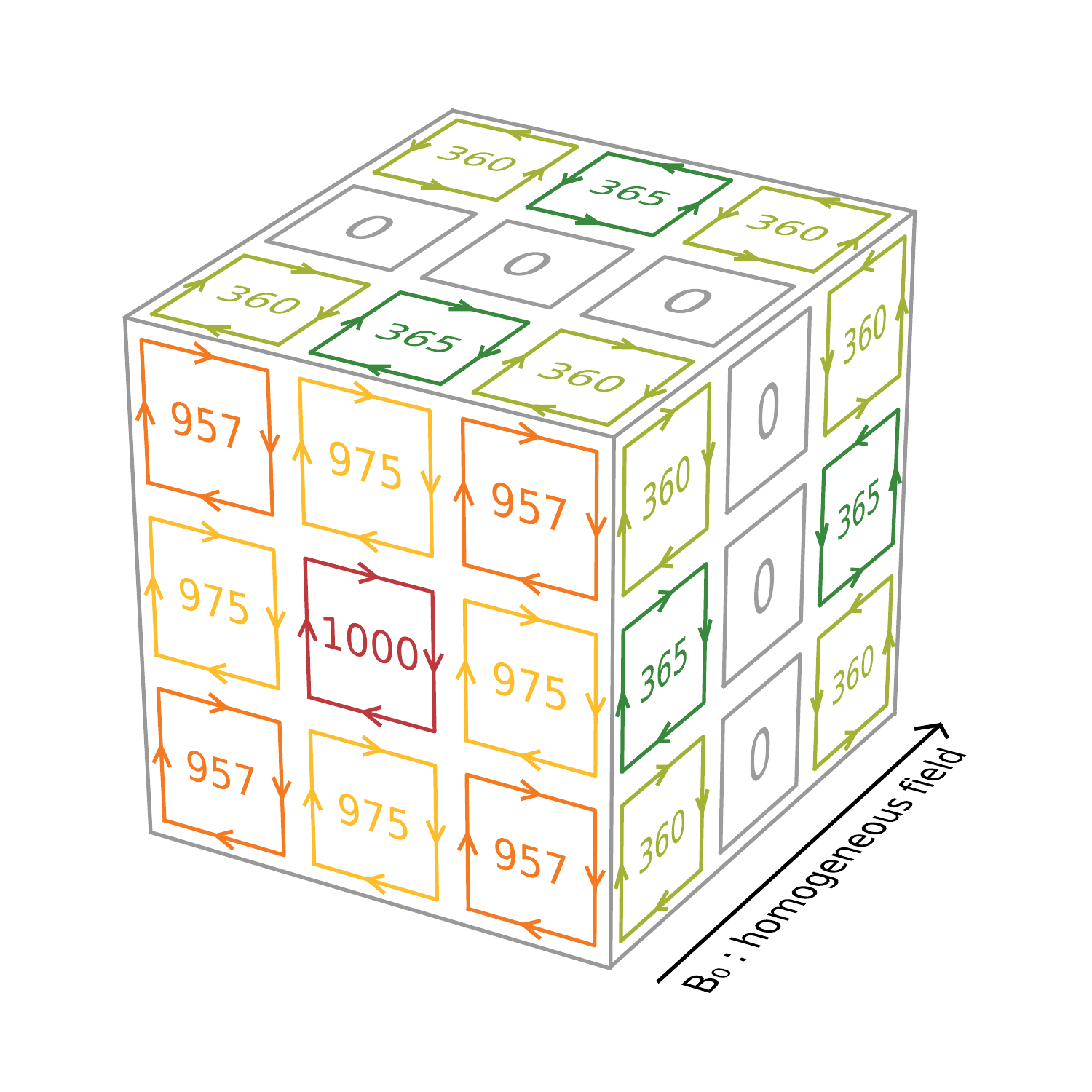}
  \caption{A solution of a tile system with $N = 6 \times (3 \times 3)$ tiles on a unit cube for a homogeneous field. The volume of interest is a cube with side length 0.75, centered inside the unit cube. Numbers indicate currents in the tile coils in arbitrary units. The currents are normalized so that the highest is 1000. For clarity, the coils are depicted slightly smaller; in the model their edges overlap. The currents on the three invisible faces are by symmetry analogous to the visible ones.}
  \label{fig:homogeneous_tiles}
\end{figure}


\begin{figure*}[bth]
  \centering
  \subfloat{
    \label{fig:homogeneous_performance_a}
    \includegraphics[width=.45\linewidth]{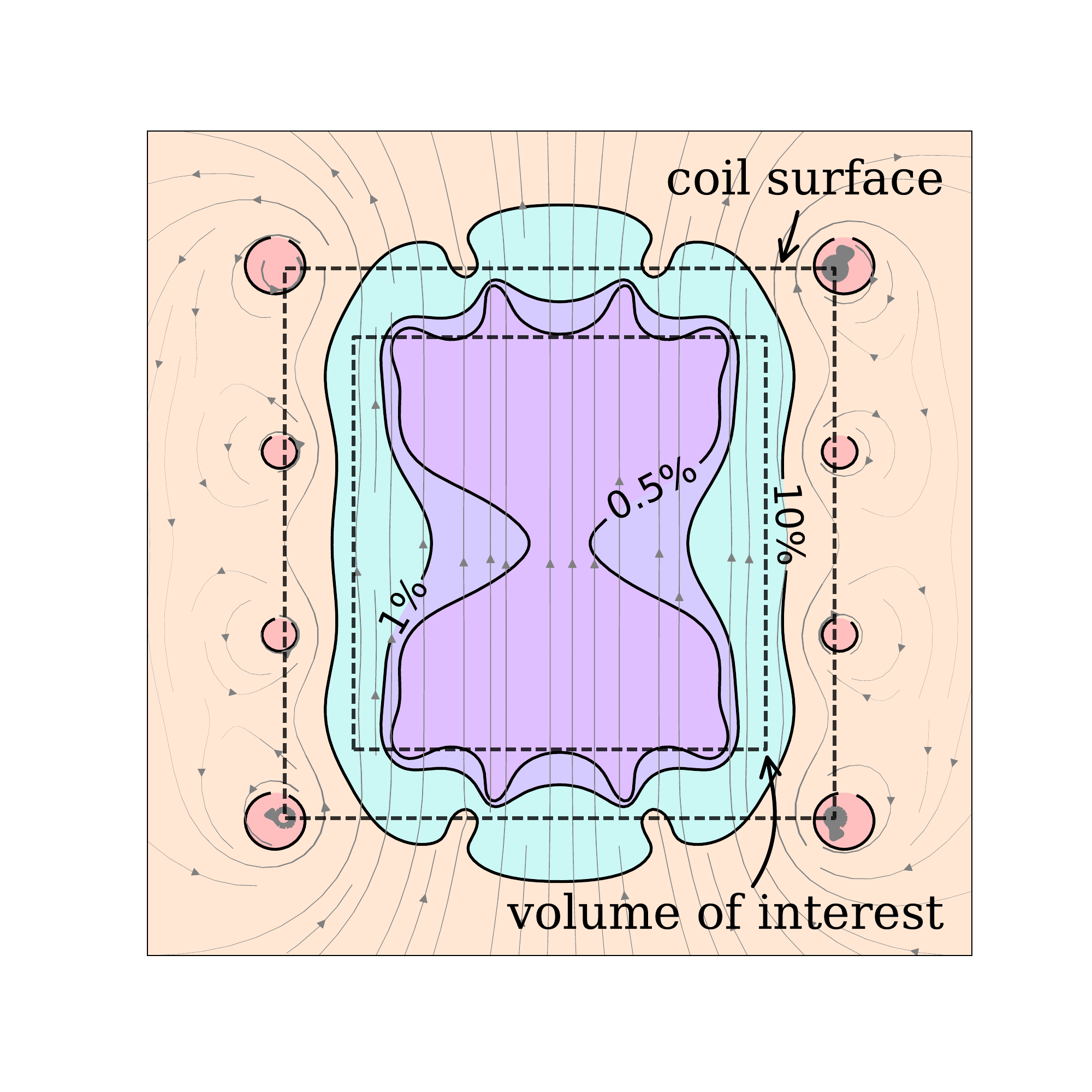}}
  \quad
  \subfloat{
    \label{fig:homogeneous_performance_b}
    \includegraphics[width=.45\linewidth]{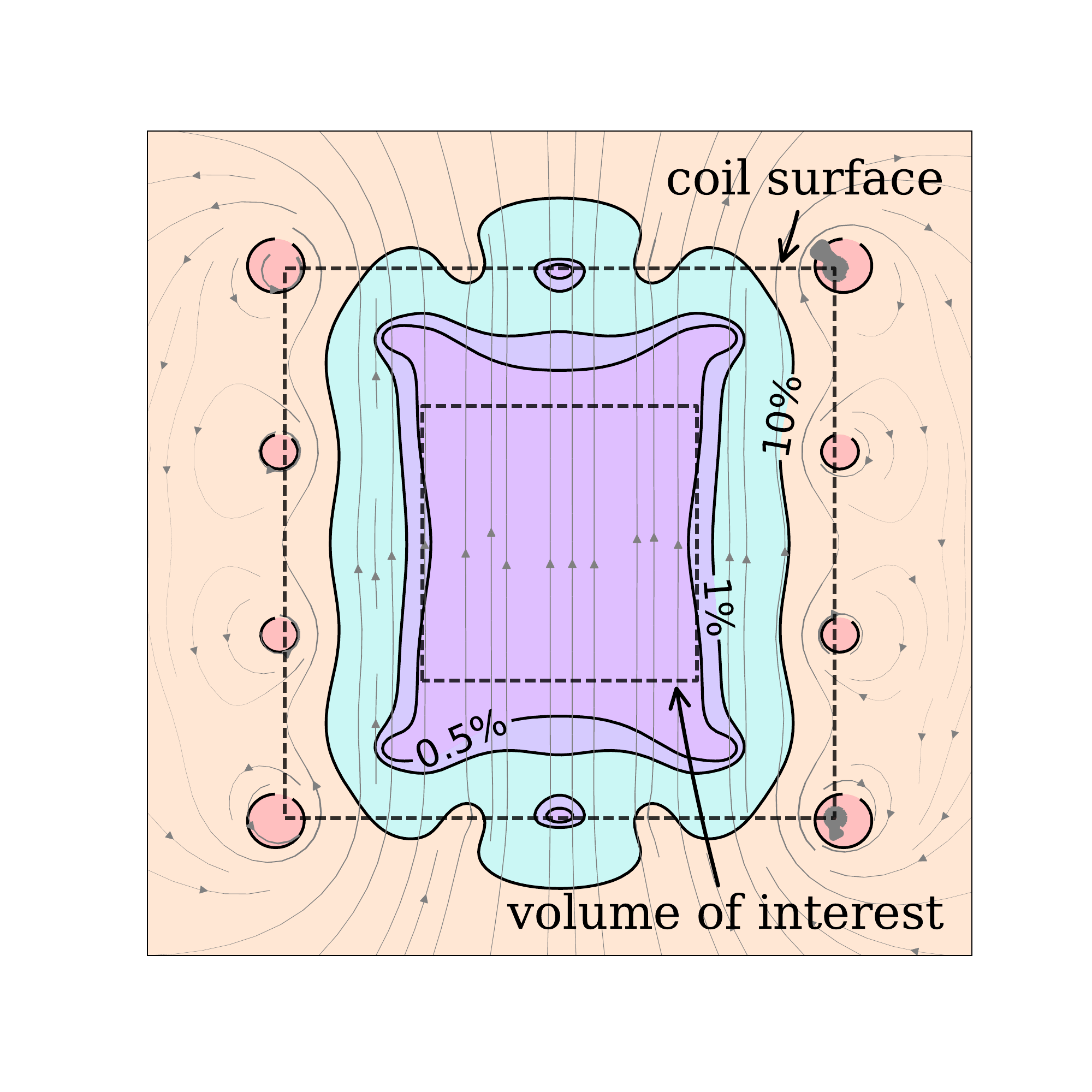}}
  \caption{Magnetic field produced by a coil designed for a homogeneous field, with $N = 6 \times (3 \times 3)$ tiles on a unit cube. The field lines are depicted in grey. Contours show boundaries of 0.5, 1 and 10\% magnitude deviation from an ideal homogeneous field. Horizontal cross sections in the middle--height plane are shown. Two designs are presented. Left-hand side: the volume of interest is a cube with side length 0.75 (the individual tile coil currents are depicted in Fig.\,\ref{fig:homogeneous_tiles}), right-hand side: the size of volume of interest is reduced to 0.5.}
  \label{fig:homogeneous_performance}
\end{figure*}

Let us look at an example of a coil design on a unit cube with the number of tiles $N = 6 \times (3 \times 3)$ (see Fig.\,\ref{fig:homogeneous_tiles}). As the volume of interest we pick a cube, centered with the unit one, with side length 0.75 (with a regular mesh of $10 \times 10$ points on each face, a total of $m = 488$ points of interest). For the sake of simplicity we design a coil for a homogeneous field along an axis of the cube. The solution of Eq.\,\ref{eq:requirement}, $\mathbb{I}_0$, directly gives the currents in each tile, which are graphically depicted in Fig.\,\ref{fig:homogeneous_tiles}. Note that many currents almost cancel each other, in particular those along horizontal edges. The magnetic field produced by the solution is shown in Fig.\,\ref{fig:homogeneous_performance_a}, as a horizontal cut along the central plane. Contours show the relative deviation from the homogeneous field. Inside the volume of interest, depicted with a dashed line, the design goal of a homogeneous field, is reproduced with few per cent accuracy. The solution, and thus the contours too, depend on the choice of the volume of interest. In general, the smaller the volume of interest, the better the accuracy. If the side length of the volume of interest is decreased to 0.5, the accuracy improves to 1\%, as shown in Fig.\,\ref{fig:homogeneous_performance_b}. Note that the optimal solution, and thereby the shape of the precision contours, change. Naturally, one can increase the accuracy also by increasing the number of tiles.


\section{Simplification of the tile system}
The tile system may find an interesting practical application. Once independently controllable tiles have been built, it can be used to produce an arbitrary field. However, building many independently driven coils is a high price to pay if one wants to produce only a simple field. Additionally, note that each edge is shared between two tiles, and the effective current is the sum of two. They may add either constructively or destructively. If the given solution is dominated by subtraction of large currents, a lot of power is unnecessarily dissipated in the system. It turns out that both problems can be solved by simplifying the tile solution.

\begin{figure*}
  \centering
  \subfloat{
    \label{fig:coils_dipole_3d_1}
    \includegraphics[width=.4\linewidth]{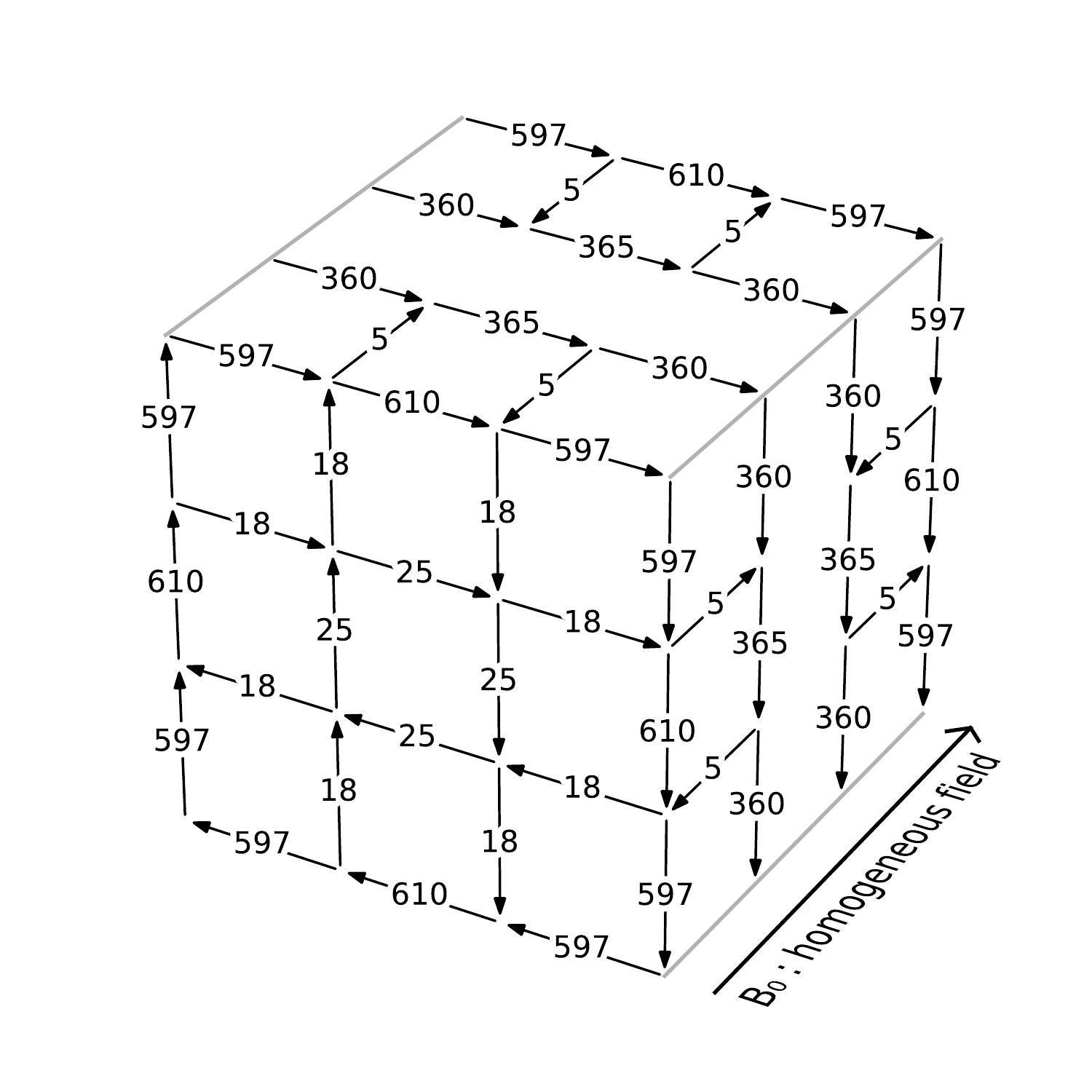}}
  \quad
  \subfloat{
    \label{fig:coils_dipole_section_0}
    \includegraphics[width=.4\linewidth]{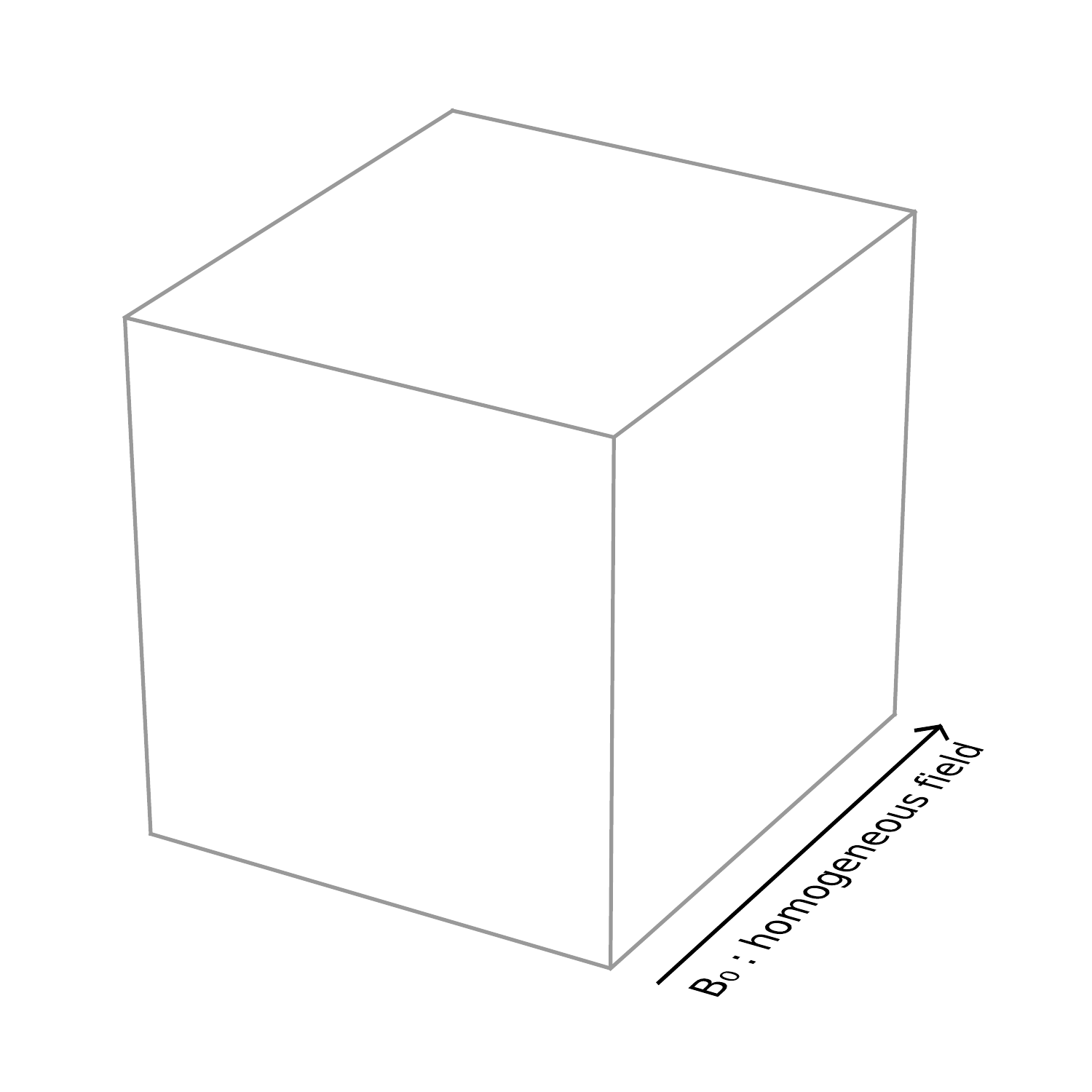}}
  \\
  \subfloat{
    \label{fig:coils_dipole_3d_3}
    \includegraphics[width=.4\linewidth]{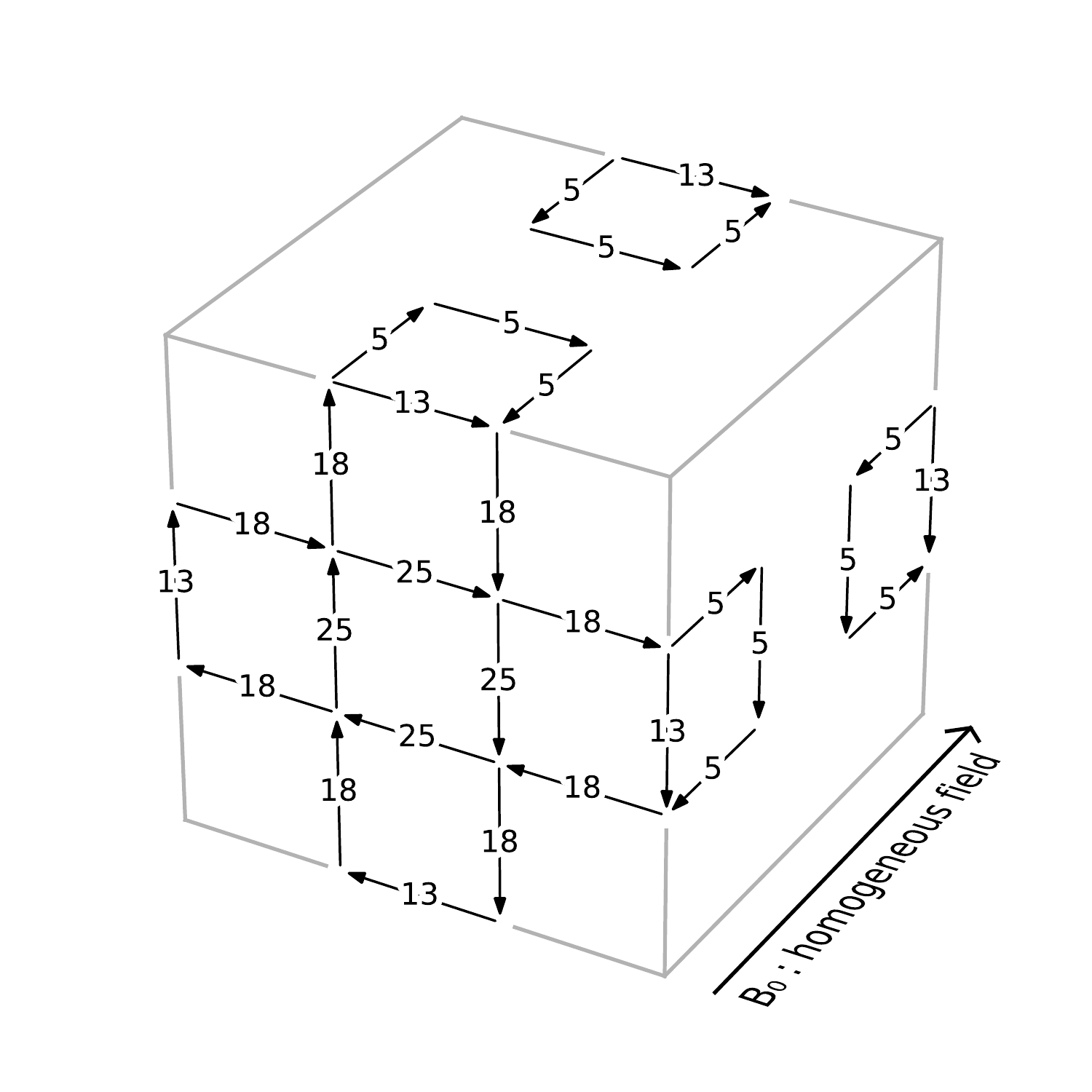}}
  \quad
  \subfloat{
    \label{fig:coils_dipole_section_2}
    \includegraphics[width=.4\linewidth]{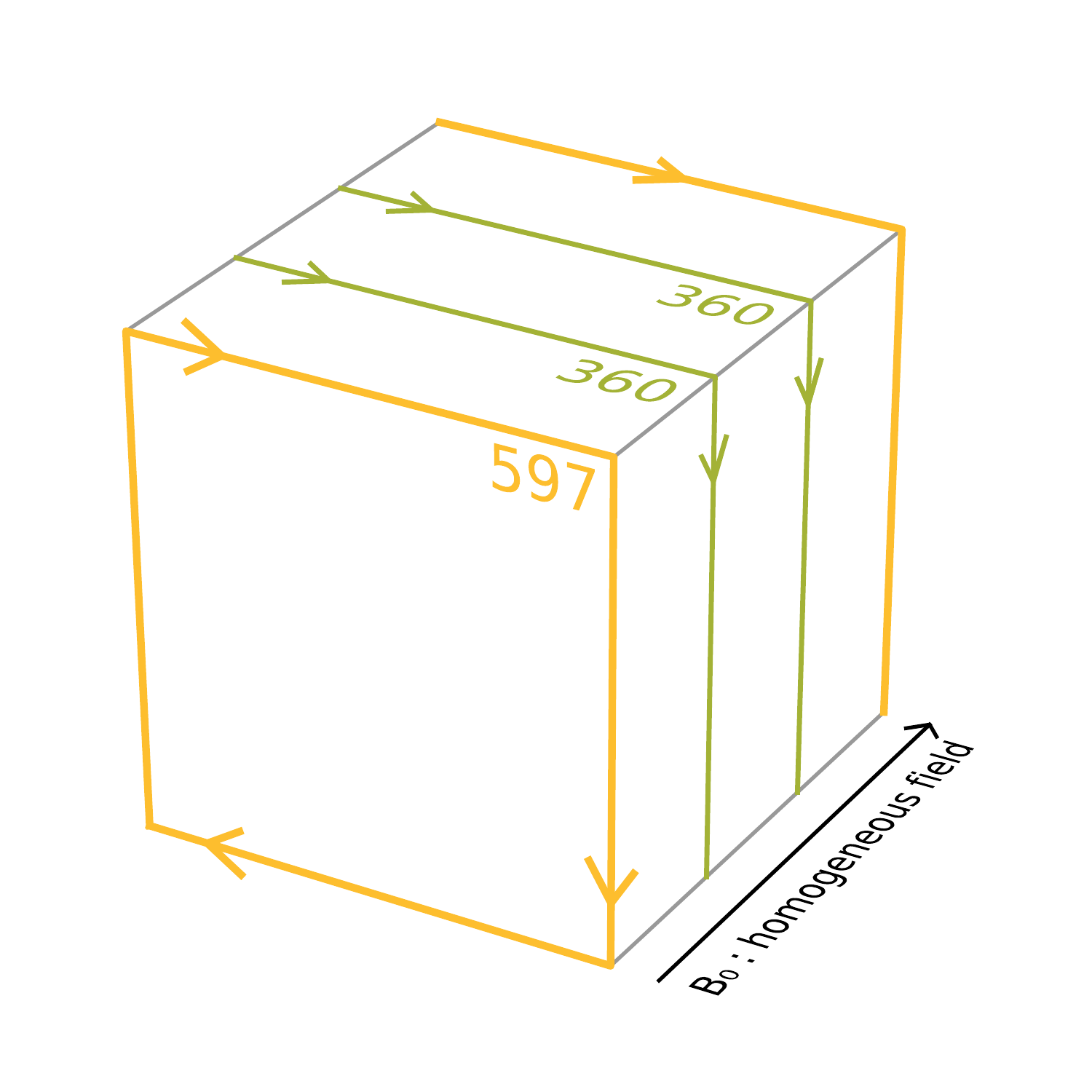}}
  \\
  \subfloat{
    \label{fig:coils_dipole_3d_5}
    \includegraphics[width=.4\linewidth]{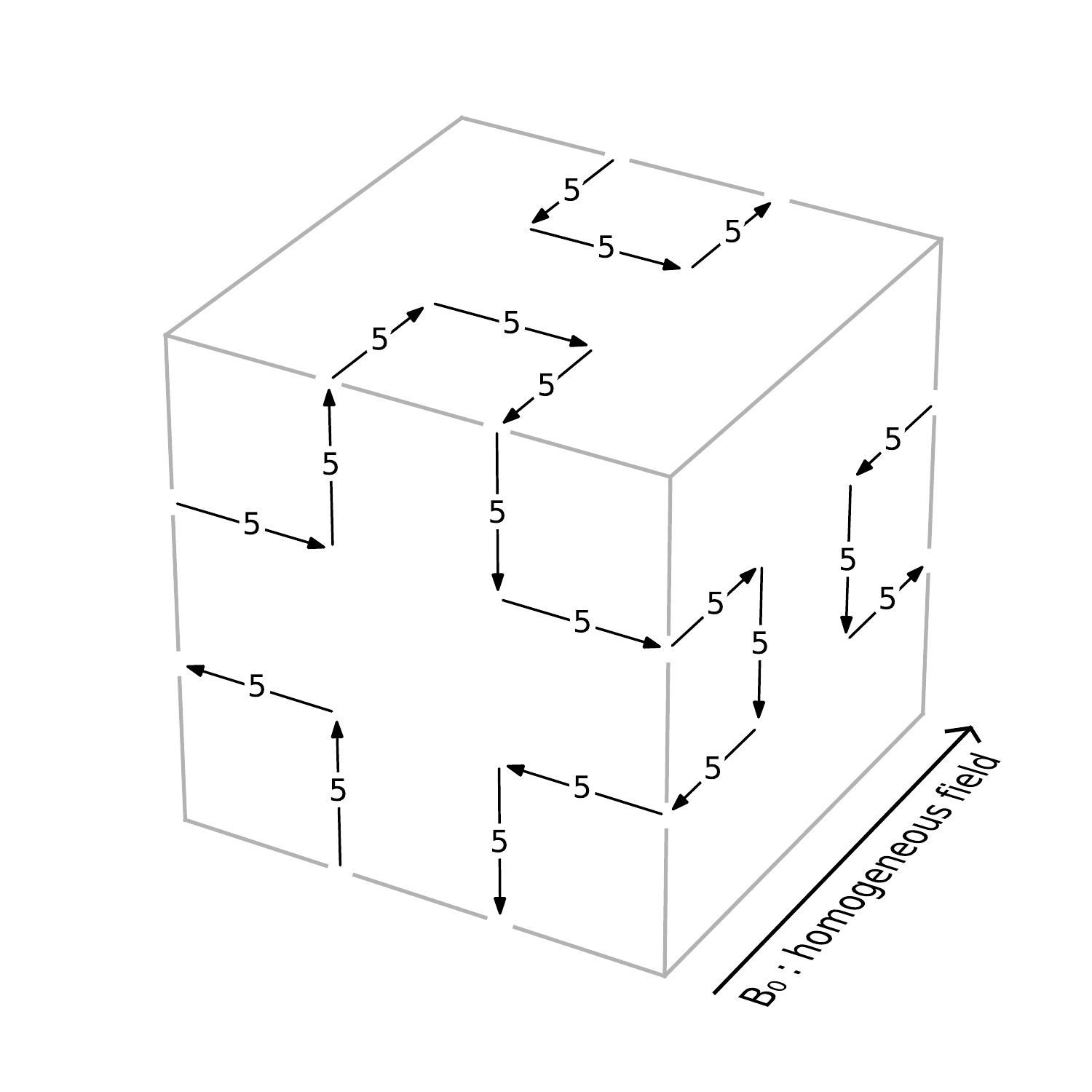}}
  \quad
  \subfloat{
    \label{fig:coils_dipole_section_4}
    \includegraphics[width=.4\linewidth]{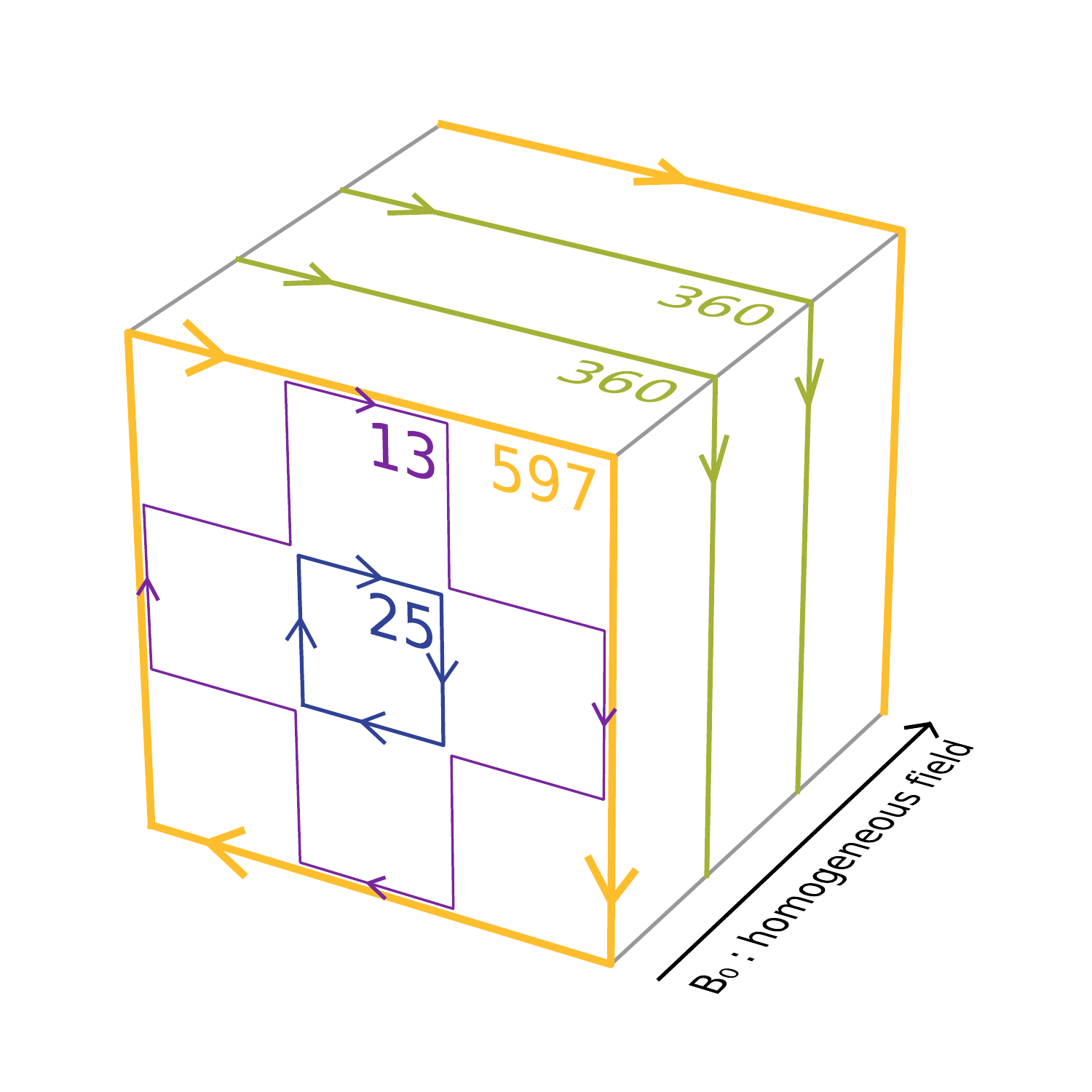}}
  \caption{Following the algorithm to simplify a coil. The left column shows the net of a current with the total current along edges of tiles. In each iteration the loop with the highest current is found and transferred onto the simplified solution, shown in the right column. We show iterations, from top: zeroth, fourth and eighth.}
  \label{fig:simplification_algorithm}
\end{figure*}

\begin{figure}
  \centering
  \includegraphics[width=\linewidth]{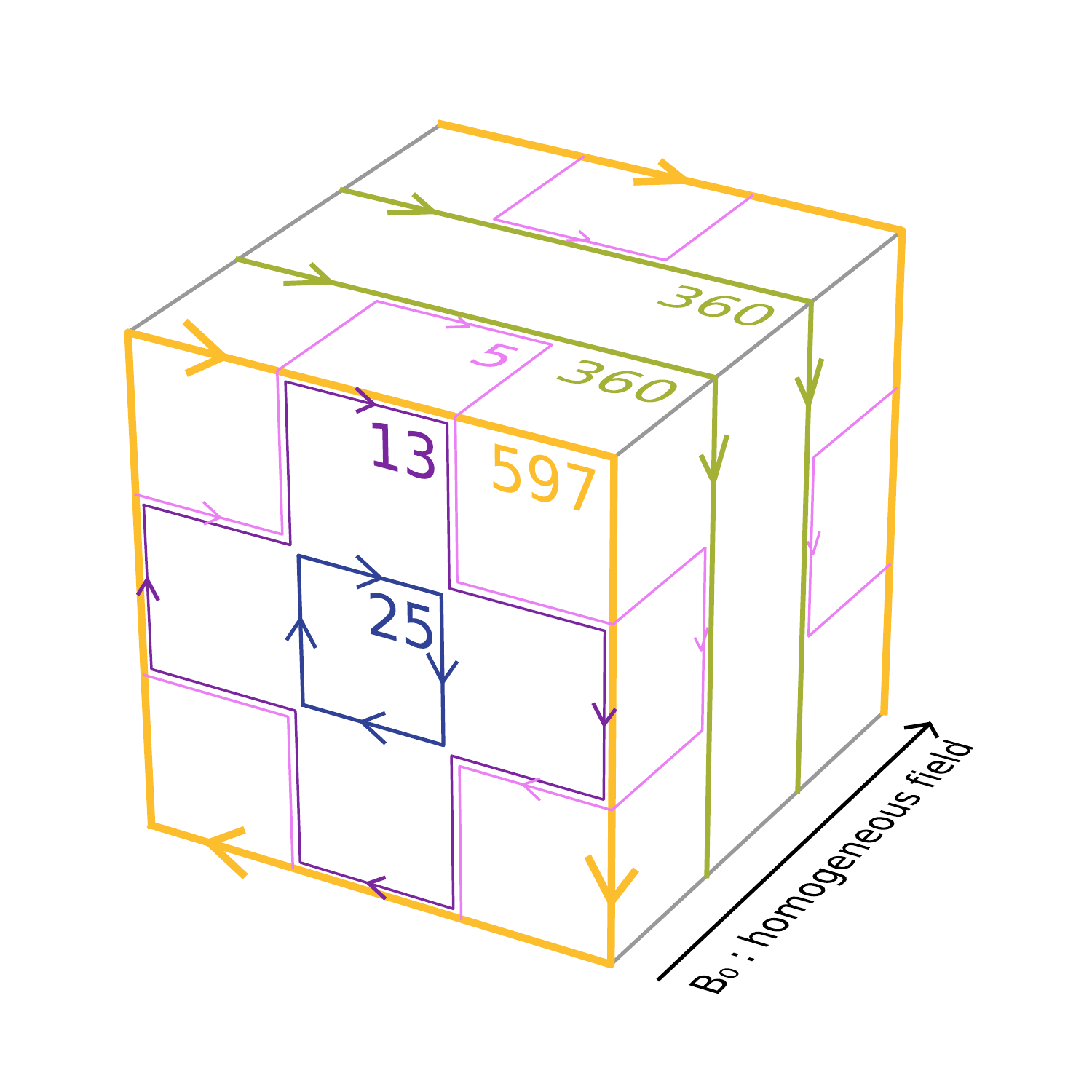}
  \caption{The coil designed for a homogeneous field, with $N = 6 \times (3 \times 3)$ tiles (Fig.\,\ref{fig:homogeneous_tiles}), simplified by adding the currents along each edge and decomposing into current loops.}
  \label{fig:homogeneous_coils}
\end{figure}

One starts by adding the currents of the adjacent tiles and assigning the sum to each common edge. The result is a complicated net of currents (upper left corner of Fig.\,\ref{fig:simplification_algorithm}). Still, each node fulfills Kirchhoff's laws. The net can then be decomposed into simple current loops by following the algorithm: Find in the net the loop with the highest current -- it will make the first one in the simplified solution. Then subtract from the net the current of the first loop along its edges. Continue to find the loop with the highest current in the modified net, which will give the second loop and continue. Some of the intermediate steps for the example (Fig.\,\ref{fig:homogeneous_tiles}) are shown in Fig.\,\ref{fig:simplification_algorithm}. The simplified solution is shown in Fig.\,\ref{fig:homogeneous_coils}. The currents in the simplified coil system are much smaller, the highest being 597 instead of 1000 and they always add constructively. Also the number of separate loops is decreased from 42 to 10. Still, the total current along each edge of a tile is exactly the same as in the tile configuration.

We conclude here our method of coil design. The simplified arrangement of coils is the optimal one, given the grid restriction, for approximating the magnetic field in the volume of interest. We continue to consider practical aspects, relevant for constructing the designs of our method.

\section{Practical considerations}
The primary practical advantage of our coil design method is that the coils are constrained to a predefined grid. This is contrary to other methods of coil design, where the position of the wires is the output of the procedure \cite{Turner1993, Beidler1990}. This may prove useful in applications with spatial constraints. Typically, coils need to be incorporated into a setup in which other components penetrate the surface on which the wires are laid. In our method it is possible to simply define the grid so that no collisions occur. Although the simple examples presented before used regular grids, we have not used symmetries to solve the problem. When many coils are designed and built, for instance to produce homogeneous magnetic fields in each of the three dimensions, they can all share the same grid. The grid can, for example, be constructed out of cable channels into which the wires are laid.

In our solution to produce the desired field one still needs a system of several coils, even in the simplified solution. The more complicated the goal field and the more tiles, the more different currents are needed across the individual loops, which quickly becomes impractical. There are several ways to tackle the problem.

The first way is to use only one current and adjust with the number of windings. In the example, when one decides for 60 as the maximum, then the current is $\mathrm{round}(597 / 60) = 10$. The 597, 360, 25, 13 and 5 would be created with 60, 36, 3, 1 and 1 windings respectively. A discretization error of $10 / 597 = 1.7\%$ is of the same order as the accuracy of the solution in representing the field (see Fig.\,\ref{fig:homogeneous_performance}). For more precise designs the numbers of windings get larger, which is troublesome to construct and causes the coils to have larger inductances.


A second way is to use a current divider. Connect the different loops in parallel, each with an appropriately chosen resistance in series. This way the ratios between the currents in each loop can be tuned precisely. However, a practical realization will most likely involve routing all loops out of the system where the current divider is installed. For more complicated coil systems with tens of different currents this may be impractical.



Yet another way is to split the loops into decades of currents. In the coil we use as an example the currents 597, 360, 13, 7, 5 (in arbitrary units) may be constructed from a set of wires with three relative currents of 100, 10 and 1, in the following way:
\begin{align*}
  597 & = 5 \times 100 + 9 \times 10 + 7 \times 1 \\
  360 & = 3 \times 100 + 6 \times 10 + 0 \times 1 \\
  13 & = 0 \times 100 + 1 \times 10 + 3 \times 1 \\
  7 & = 0 \times 100 + 0 \times 10 + 7 \times 1 \\
  5 & = 0 \times 100 + 0 \times 10 + 5 \times 1 \\
\end{align*}
In this way one can reach better than 1\% accuracy in reproducing the solution in practice with only 3 different currents to control, even for complicated designs. Those can be either separately controlled or split with a current divider.

We do not consider any of the above ways superior. It is up to the particular application which is the best suited one.




\section{An example of an application}



The presented method may prove useful to precision physics experiments. Among them is the measurement of the electric dipole moment of the neutron, an observable providing direct insight into the fundamental $\mathcal{CP}$ symmetry breaking in the strong interaction \cite{khriplovic:1997}. In the experiment at the Paul Scherrer Institute the electric dipole moment of the neutron is measured with a precise spectroscopy of polarized neutrons in a combination of electric and magnetic fields \cite{Baker2011}. The precision of the measurement sets very strict requirements for the magnetic field stability, yet the experiment is only one of the many in the experimental hall. Some of the other setups use strong magnets, often ramped up and down daily. The neutron electric dipole moment measurement would not be possible without counter measures against the external magnetic field changes.

The experiment uses an active magnetic field stabilization system. The apparatus is located in the middle of a set of large coils, connected in a feedback loop with magnetic field sensors. The present system is described in detail in \cite{Afach2014}. The geometry of the coils, six in total, is very simple: each coil is a rectangle and they are all arranged in three perpendicular Helmholtz--like pairs. The simple geometry causes the volume in which the field is stabilized (the volume of interest) to be small relative to the size of the coils. Only thanks to the sufficient space around the less than 3$\,$m large apparatus it was possible to make these simple coils large enough, 6--8$\,$ meters side length, for the experiment to fit in the volume of interest. Also, the variety of field shapes that can be compensated is limited to homogeneous ones (parallel currents in a Helmholtz--like pair) plus fields created when each coil is controlled separately.

Using the method presented here to design the active compensation coils offers improvements in two areas. Firstly, the size of the coils could be decreased, or the size of the experimental set-up increased, without loss of performance. Secondly, more coils could be built, extending the range of possible fields to be compensated. These could either be generic: e.g. pure independent gradients, or coils dedicated to counteract a particular known disturbance, designed based on a field map as the goal field in the described method.


\begin{figure*}
  \centering
  \subfloat{
    \label{fig:coils_dipole_3d}
    \includegraphics[width=.35\linewidth]{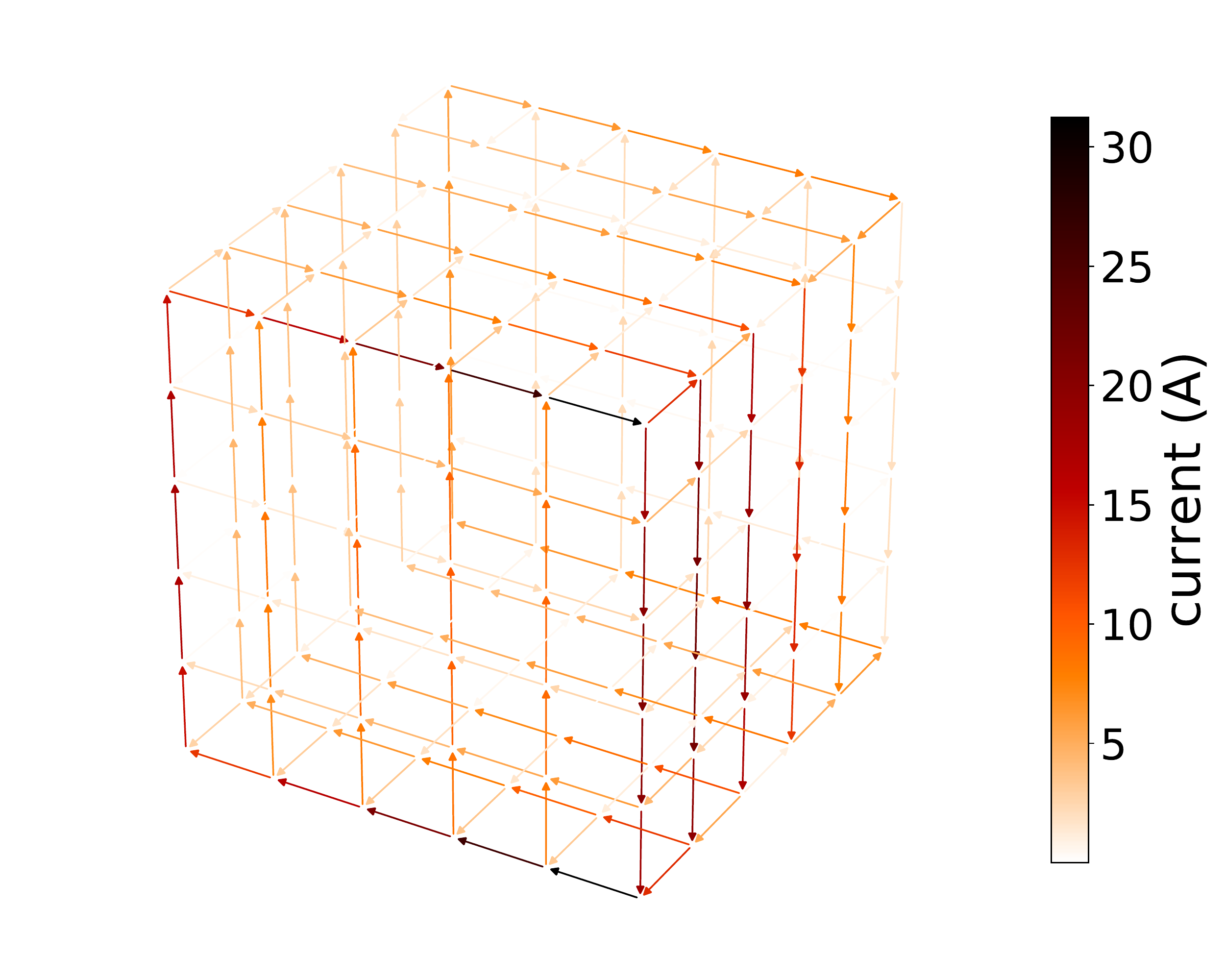}}
  \quad
  \subfloat{
    \label{fig:coils_dipole_section}
    \includegraphics[width=.55\linewidth]{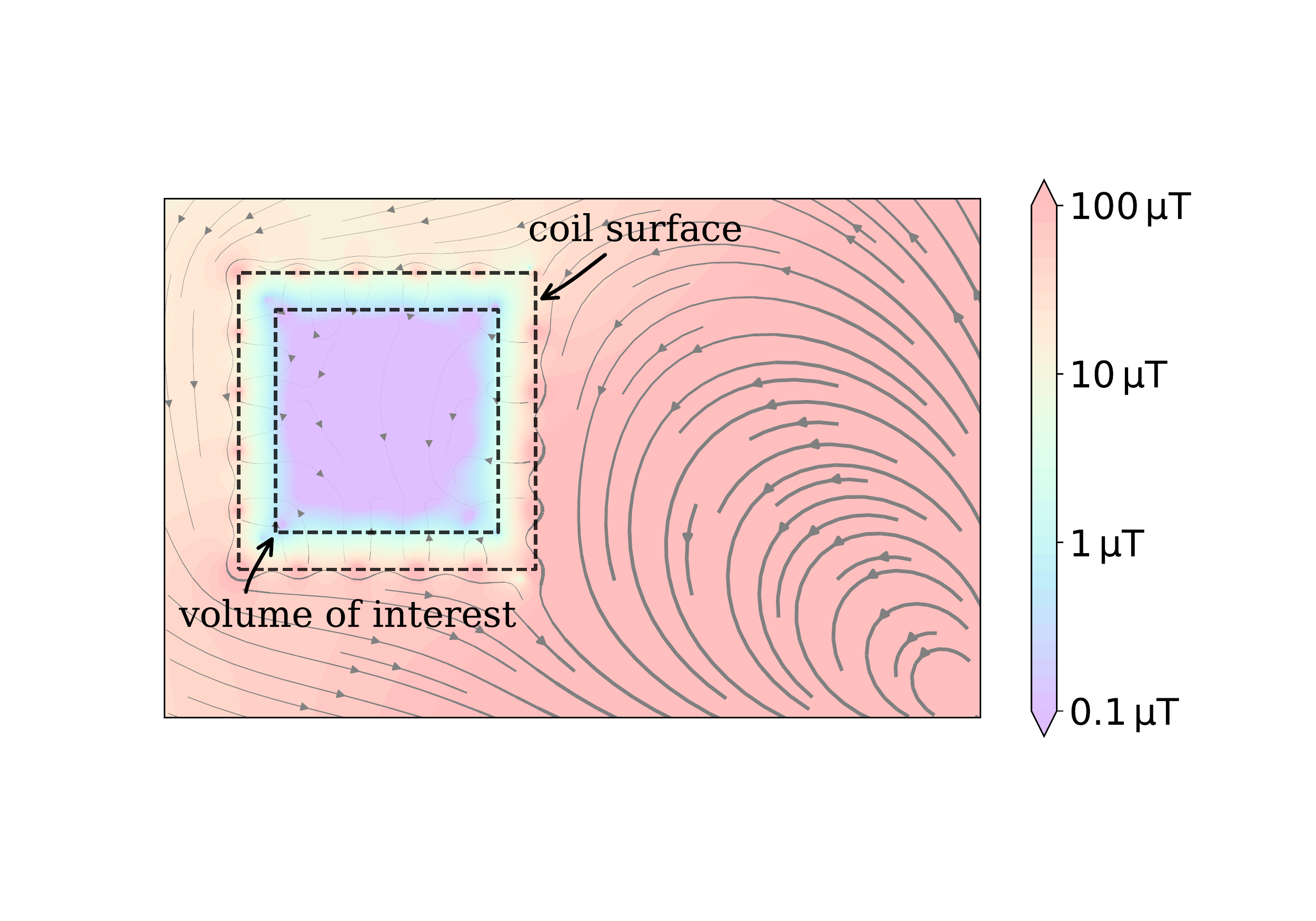}}
  \caption{A coil designed on a unit cube with $5 \times 5$ tiles per face to shield against a dipole disturbance. The dipole is located, relative to the center of the unit cube, two units to the right and one unit to the front. It is located in the middle height of the cube. The volume of interest has a side length of 0.75. On the left--hand side the total current along each edge of the dipole compensation coil is depicted. On the right--hand side the magnetic field is shown. The magnetic field lines are shown in grey, the volume of interest and the coil surface with dashed lines. The colors depict the magnitude of the magnetic field (capped at 0.1 and 100\,µT). A horizontal cross section in the middle height is shown. The dipole source is located in the lower right corner of the plot and points parallel to the plane of the plot. The magnitude of the field in the volume of interest is reduced from tens of microteslas down to below one.}
  \label{fig:showcase}
\end{figure*}

\section{Conclusion}
Coil design is a complicated and very technical problem, especially when high accuracy is required. We presented a method that is simple in terms of both underlying math and computational effort. We believe that the design method can find its niche in practical applications, where spatial constraints play a significant role and a percent level in accuracy of the produced field is acceptable.

For the sake of clarity we explained the method on a simple example. We designed a coil for a homogeneous field with only few tiles. However, the method is much more powerful. We conclude by presenting, in Fig.\,\ref{fig:showcase}, a showcase design with $N = 6 \times (5 \times 5) = 150$ tiles for compensating a nearby dipole source.

We publish the software implementation of the coil design, including examples, as open--source \cite{Coilsjlcode}.

We would like to thank Georg Bison, Allard Schnabel and other members of the nEDM at PSI collaboration for fruitful discussions. This work was supported by the Swiss National Science Foundation under grant 200020\_162574.

\bibliography{Bibliography}

\begin{thebibliography}{7}%
\makeatletter
\providecommand \@ifxundefined [1]{%
 \@ifx{#1\undefined}
}%
\providecommand \@ifnum [1]{%
 \ifnum #1\expandafter \@firstoftwo
 \else \expandafter \@secondoftwo
 \fi
}%
\providecommand \@ifx [1]{%
 \ifx #1\expandafter \@firstoftwo
 \else \expandafter \@secondoftwo
 \fi
}%
\providecommand \natexlab [1]{#1}%
\providecommand \enquote  [1]{``#1''}%
\providecommand \bibnamefont  [1]{#1}%
\providecommand \bibfnamefont [1]{#1}%
\providecommand \citenamefont [1]{#1}%
\providecommand \href@noop [0]{\@secondoftwo}%
\providecommand \href [0]{\begingroup \@sanitize@url \@href}%
\providecommand \@href[1]{\@@startlink{#1}\@@href}%
\providecommand \@@href[1]{\endgroup#1\@@endlink}%
\providecommand \@sanitize@url [0]{\catcode `\\12\catcode `\$12\catcode
  `\&12\catcode `\#12\catcode `\^12\catcode `\_12\catcode `\%12\relax}%
\providecommand \@@startlink[1]{}%
\providecommand \@@endlink[0]{}%
\providecommand \url  [0]{\begingroup\@sanitize@url \@url }%
\providecommand \@url [1]{\endgroup\@href {#1}{\urlprefix }}%
\providecommand \urlprefix  [0]{URL }%
\providecommand \Eprint [0]{\href }%
\providecommand \doibase [0]{http://dx.doi.org/}%
\providecommand \selectlanguage [0]{\@gobble}%
\providecommand \bibinfo  [0]{\@secondoftwo}%
\providecommand \bibfield  [0]{\@secondoftwo}%
\providecommand \translation [1]{[#1]}%
\providecommand \BibitemOpen [0]{}%
\providecommand \bibitemStop [0]{}%
\providecommand \bibitemNoStop [0]{.\EOS\space}%
\providecommand \EOS [0]{\spacefactor3000\relax}%
\providecommand \BibitemShut  [1]{\csname bibitem#1\endcsname}%
\let\auto@bib@innerbib\@empty
\bibitem [{\citenamefont {Turner}(1993)}]{Turner1993}%
  \BibitemOpen
  \bibfield  {author} {\bibinfo {author} {\bibfnamefont {R.}~\bibnamefont
  {Turner}},\ }\href {\doibase 10.1016/0730-725X(93)90209-V} {\bibfield
  {journal} {\bibinfo  {journal} {Magn Reson Imaging}\ }\textbf {\bibinfo
  {volume} {11}},\ \bibinfo {pages} {903} (\bibinfo {year} {1993})}\BibitemShut
  {NoStop}%
\bibitem [{\citenamefont {Beidler}\ \emph {et~al.}(1990)\citenamefont {Beidler}
  \emph {et~al.}}]{Beidler1990}%
  \BibitemOpen
  \bibfield  {author} {\bibinfo {author} {\bibfnamefont {C.}~\bibnamefont
  {Beidler}} \emph {et~al.},\ }\href {\doibase
  http://www.ans.org/pubs/journals/fst/a_29178} {\bibfield  {journal} {\bibinfo
   {journal} {Fusion Technology}\ }\textbf {\bibinfo {volume} {17}},\ \bibinfo
  {pages} {148} (\bibinfo {year} {1990})}\BibitemShut {NoStop}%
\bibitem [{\citenamefont {Afach}\ \emph {et~al.}(2014)\citenamefont {Afach}
  \emph {et~al.}}]{Afach2014}%
  \BibitemOpen
  \bibfield  {author} {\bibinfo {author} {\bibfnamefont {S.}~\bibnamefont
  {Afach}} \emph {et~al.},\ }\href {\doibase 10.1063/1.4894158} {\bibfield
  {journal} {\bibinfo  {journal} {Journal of Applied Physics}\ }\textbf
  {\bibinfo {volume} {116}},\ \bibinfo {pages} {084510} (\bibinfo {year}
  {2014})}\BibitemShut {NoStop}%
\bibitem [{\citenamefont {Reta-Hernandez}\ and\ \citenamefont
  {Karady}(1998)}]{Reta-Hernandez1998}%
  \BibitemOpen
  \bibfield  {author} {\bibinfo {author} {\bibfnamefont {M.}~\bibnamefont
  {Reta-Hernandez}}\ and\ \bibinfo {author} {\bibfnamefont {G.~G.}\
  \bibnamefont {Karady}},\ }\href {\doibase Doi 10.1016/S0378-7796(97)01232-7}
  {\bibfield  {journal} {\bibinfo  {journal} {Electric Power Systems Research}\
  }\textbf {\bibinfo {volume} {45}},\ \bibinfo {pages} {57} (\bibinfo {year}
  {1998})}\BibitemShut {NoStop}%
\bibitem [{\citenamefont {Khriplovich}\ and\ \citenamefont
  {Lamoreaux}(1997)}]{khriplovic:1997}%
  \BibitemOpen
  \bibfield  {author} {\bibinfo {author} {\bibfnamefont {I.}~\bibnamefont
  {Khriplovich}}\ and\ \bibinfo {author} {\bibfnamefont {S.}~\bibnamefont
  {Lamoreaux}},\ }\href@noop {} {\emph {\bibinfo {title} {{C}P {V}iolation
  {W}ithout {S}trangeness}}}\ (\bibinfo  {publisher} {Springer},\ \bibinfo
  {year} {1997})\BibitemShut {NoStop}%
\bibitem [{\citenamefont {Baker}\ \emph {et~al.}(2011)\citenamefont {Baker}
  \emph {et~al.}}]{Baker2011}%
  \BibitemOpen
  \bibfield  {author} {\bibinfo {author} {\bibfnamefont {C.~A.}\ \bibnamefont
  {Baker}} \emph {et~al.},\ }\href {\doibase 10.1016/j.phpro.2011.06.032}
  {\bibfield  {journal} {\bibinfo  {journal} {Physics Procedia}\ }\textbf
  {\bibinfo {volume} {17}},\ \bibinfo {pages} {159} (\bibinfo {year}
  {2011})}\BibitemShut {NoStop}%
\bibitem [{\citenamefont {Rawlik}()}]{Coilsjlcode}%
  \BibitemOpen
  \bibfield  {author} {\bibinfo {author} {\bibfnamefont {M.}~\bibnamefont
  {Rawlik}},\ }\href {\doibase 10.5905/ethz-1007-110} {}\bibinfo {note}
  {{\url{github.com/rawlik/Coils.jl}, or
  \url{doi.org/10.5905/ethz-1007-110}}}\BibitemShut {NoStop}%
\end{thebibliography}%

\end{document}